\documentclass[12pt,preprint]{aastex}

\def\kms{\ifmmode{\rm km\thinspace s^{-1}}\else km\thinspace s$^{-1}$\fi}
\def\ms{\ifmmode{\rm m\thinspace s^{-1}}\else m\thinspace s$^{-1}$\fi}

\usepackage{amsmath,amsfonts}

\begin{document}

\title{New Data and Improved Parameters for the Transiting Planet
OGLE-TR-56b}

\author{Guillermo Torres\altaffilmark{1}, Maciej Konacki\altaffilmark{2},
Dimitar D. Sasselov\altaffilmark{3} and Saurabh Jha\altaffilmark{4}}

\altaffiltext{1}{Harvard-Smithsonian Center for Astrophysics, 60 Garden St.,
Cambridge, MA 02138, USA; e-mail: gtorres@cfa.harvard.edu}
\altaffiltext{2}{Department of Geological and Planetary Sciences, California
Institute of Technology, MS 150-21, Pasadena, CA 91125, USA;
Nicolaus Copernicus Astronomical Center, Polish Academy of Sciences,
Rabia\'nska 8, 87-100 Toru\'n, Poland; e-mail: maciej@gps.caltech.edu}
\altaffiltext{3}{Harvard-Smithsonian Center for Astrophysics, 60 Garden St.,
Cambridge, MA 02138, USA; e-mail: dsasselov@cfa.harvard.edu}
\altaffiltext{4}{Department of Astronomy, University of California,
Berkeley, CA 94720, USA; e-mail: saurabh@astron.berkeley.edu}

\begin{abstract} 

We report new spectroscopic observations of the recently discovered
transiting planet OGLE-TR-56b with the Keck/HIRES instrument. Our
radial velocity measurements with errors of $\sim$100~$\ms$ show clear
variations that are in excellent agreement with the phasing (period
and epoch) derived from the OGLE transit photometry, confirming the
planetary nature of the companion. The new data combined with
measurements from the previous season allow an improved determination
of the mass of the planet, $M_p = 1.45 \pm 0.23~M_{\rm Jup}$. All
available OGLE photometry, including new measurements made this
season, have also been analyzed to derive an improved value for the
planetary radius of $R_p = 1.23 \pm 0.16~R_{\rm Jup}$. We discuss the
implications of these results for the theory of extrasolar planets.

\end{abstract}

\keywords{techniques: radial velocities --- binaries: eclipsing ---
stars: low-mass, brown dwarfs --- planetary systems}

\section{Introduction}
\label{sec:introduction}

Most extrasolar planets to date have been discovered with the
high-precision radial velocity technique, which provides only a lower
limit to the mass of the companion because the inclination angle
cannot be determined from spectroscopy alone.  Systems for which the
orbit happens to be nearly edge-on, so that the planet transits across
the disk of the star once every orbital period, show a photometric
transit and allow the absolute mass of the planet to be determined.
Transiting systems are valuable in many other ways, providing the
planet's absolute radius, as well as allowing a variety of different
follow-up studies \citep[see, e.g.,][]{Brown:02, Charbonneau:02,
Vidal:03, Fortney:03, Richardson:03, Moutou:03}. Transits are also a
viable planet discovery technique: our recent follow-up in 2002 of
candidates from the OGLE-III sample toward the bulge of the Galaxy
\citep{Udalski:02a,Udalski:02b} resulted in the spectroscopic
confirmation of a planet around the star OGLE-TR-56 ($V = 16.6$), with
a period of 1.2~days. This is the first case originally discovered
from its photometric signature rather than its Doppler signature
\citep{Konacki:03a}.

The limited amount of spectroscopic data we obtained during our 2002
season only allowed for a relatively uncertain estimate of the mass of
OGLE-TR-56b. A combined orbital solution using our velocities and the
OGLE-III light curve yielded $M_p = 0.9 \pm 0.3$~M$_{\rm Jup}$
\citep{Konacki:03a}.  In this Letter we report new radial velocity
measurements that allow us to improve the accuracy of the mass
determination and to better characterize its uncertainty, as well as
to strengthen the case against any false-positive scenarios. In
addition, we present an updated transit light curve solution based on
improvements in the OGLE photometry.

\section{Observations and reductions}
\label{sec:observations}

OGLE-TR-56 was observed spectroscopically on 5 nights in August 2003
with the Keck~I telescope and the HIRES instrument \citep{Vogt:94}. We
obtained a total of 8 new spectra of the object, with exposure times
ranging from 30 to 50 minutes.  The setup allowed us to record 35
usable echelle orders covering the spectral range from 3850~\AA\ to
6200~\AA\ at a resolving power of $R \simeq 65,\!000$.  Typical
signal-to-noise ratios are in the range of 10--20 per pixel for a
single exposure.  Our main wavelength reference was provided by a
hollow-cathode Thorium-Argon lamp, of which we obtained short
exposures immediately preceding and following each stellar exposure. 

In addition to our program star we obtained frequent observations of
two brighter stars (HD~209458 and HD~179949) that have known
low-amplitude velocity variations at the level of about 200~$\ms$
(peak to peak) due to orbiting substellar companions \citep{Henry:00,
Charbonneau:00, Tinney:01}, and which we used as ``standards''.  These
stars were observed with the iodine gas absorption cell
\citep{Marcy:92}.  All HIRES spectra were bias-subtracted,
flat-fielded, cleaned of cosmic rays, and extracted using the MAKEE
reduction package written by Tim Barlow \citeyearpar{Barlow:02}.
Compared to the procedures followed in \cite{Konacki:03a}, a number of
details in the reductions were fine-tuned for the new observations and
led to slightly improved noise levels and better velocities.  We
therefore re-reduced the original 2002 spectra along with the new ones
for uniformity.  Wavelength solutions based on the Th-Ar exposures
were carried out with standard tasks in IRAF\footnote{IRAF is
distributed by the National Optical Astronomy Observatories, which is
operated by the Association of Universities for Research in Astronomy,
Inc., under contract with the National Science Foundation.}.

Radial velocities for OGLE-TR-56 and for the standards stars were
derived by cross-correlation against a synthetic template computed
specifically for the parameters of each star as detailed by
\cite{Konacki:03b}. For the cross-correlations we used the IRAF task
XCSAO \citep{Kurtz:98}.  The final velocities are the weighted average
of all echelle orders in each spectrum (only orders not affected by
the iodine were used for the standards).  Formal errors were derived
from the scatter of the velocities determined from the different
orders. These are typically well under $\sim$100~$\ms$, and do not
include systematic components, which we have previously estimated to
be no larger than about 100~$\ms$ for this instrumentation
\citep[see][]{Konacki:03b}. The radial velocities in the frame of the
solar system barycenter from all of the spectra (2002 and 2003) along
with their final errors are listed in Table~\ref{tab:rvs}.

\section{Spectroscopic orbital solution}
\label{sec:orbit}

The new radial velocities for OGLE-TR-56 show clear changes with
orbital phase. The latter is well known from the photometric
observations that yield a very accurate period and transit epoch (see
below). However, there is also a systematic shift compared to the 2002
velocity measurements of about 200~$\ms$. A similar shift is observed
in the standards, indicating it is a real effect. Such offsets from
run to run are common in radial-velocity work, and can be due to a
number of reasons including temperature changes and other instrumental
effects beyond the control of the observer.  In order to optimally
remove this shift using all of the available information, we developed
a procedure by which we fit for the orbits of the three stars
simultaneously. We solve for the shift at same time as the rest of the
orbital elements and assume that the offset is identical for the three
stars. The phase and velocity amplitudes of the circular orbits for
HD~209458 and HD~179949 are known from high-precision velocity work
\citep{Mazeh:00, Tinney:01}, and were held fixed.  Therefore, the five
free parameters in the least-squares problem are the semi-amplitude of
the velocity curve of OGLE-TR-56, the center-of-mass velocity for each
star, and the common offset between the 2002 and 2003 seasons. The
ephemeris for OGLE-TR-56 is also fixed, as mentioned above, to the
value determined in our light curve analysis described in
\S\ref{sec:lightcurve}.

The solution, based on a total of 28 observations (11 of our target, 9
of HD~179949, and 8 of HD~209458), gives a velocity semi-amplitude for
OGLE-TR-56 of $K = 265 \pm 38$~$\ms$. The offset between the two
observing seasons is determined to be $\Delta_{2003-2002} = +192 \pm
47$~$\ms$, and the overall RMS residual from the fit for OGLE-TR-56 is
114 $\ms$. The minimum mass for the planet in orbit around our target
is $M_p \sin i = 1.33 \pm 0.21 \times 10^{-3} \times
(M_s+M_p)^{2/3}$~M$_{\sun}$, where $M_s$ is the mass of the primary
star. The observations for OGLE-TR-56 along with the orbital fit are
shown in Figure~\ref{fig:orbit}. The measurements listed in
Table~\ref{tab:rvs} include the offset $\Delta_{2003-2002}$, so that
all measurements are referred to the 2002 frame. 

The center-of-mass velocities derived for the three stars are $-24.579
\pm 0.045~\kms$ (HD~179949), $-14.577 \pm 0.048~\kms$ (HD~209458), and
$-48.317 \pm 0.045~\kms$ (OGLE-TR-56). For the latter object the
difference compared to the value of $-49.49~\kms$ by
\cite{Konacki:03a} is due to differences in the reduction of the
spectra (\S\ref{sec:observations}) and the increased number of
observations in the present solution.  The above center-of-mass
velocities are on the reference frame of the templates used for the
cross-correlations, which are calculated spectra. The errors given are
strictly internal, and do not include contributions from uncertainties
in the instrumental zero point (of the kind that lead to
$\Delta_{2003-2002}$), or in the wavelength scale or other details of
the model atmospheres that go into the calculation of the templates.
The absolute accuracy of these velocities may be in error by several
hundred $\ms$. Nevertheless, it may be of interest for future studies
to refer the center-of-mass velocity of OGLE-TR-56 to some
well-defined frame of reference.  A comparison of our values for the
two standards against the results by \cite{Nidever:02} gives
systematic differences of $0.083~\kms$ (HD~179949) and $0.182~\kms$
(HD~209458), in the sense that our velocities are larger in both
cases. The average offset is $0.132~\kms$. Applying this correction to
OGLE-TR-56 gives the value $-48.449~\kms$ for its center-of-mass
velocity, on the same scale as \cite{Nidever:02}, with an estimated
total uncertainty of approximately 100~$\ms$. 
	
\section{Spectral line bisectors}
\label{sec:bisectors}

Following \cite{Konacki:03b} we used our new spectroscopic
observations to re-examine the possibility that the velocity
variations we measured for OGLE-TR-56 are not produced by a planet
orbiting the star, but are instead the result of a blend scenario. In
this case, small asymmetries in the spectral lines due to the presence
of another star (e.g., the primary of an eclipsing binary in the
background) can lead to spurious velocities as the second set of lines
moves back and forth in phase with the photometric period. We
investigated this for each of our spectra by computing the line
bisectors directly from the correlation functions (co-added over all
orders), which are representative of the average line profile for the
star. We then calculated the ``bisector span'' as the velocity
difference between the bisectors at two different correlation levels.
This can be used as a measure of the asymmetry of the lines
\citep[see, e.g.,][]{Santos:02}. 

In Figure~\ref{fig:bisectors} we show the bisector span for each of
our spectra as a function of orbital phase.  There is no significant
correlation with phase, supporting the conclusion that the velocity
variations we measured for the star are real. 
	
\section{Analysis of the light curve}
\label{sec:lightcurve}

Photometric observations of OGLE-TR-56 by the OGLE team have continued
after its discovery in 2001, and now include 3 observing seasons (1113
measurements covering more than 600 cycles of the orbit). A total of
13 transits have been recorded. Additionally, small corrections for
systematic errors in the photometry have recently been applied that
improve the errors slightly\footnote{See \tt
http://bulge.princeton.edu/$\sim$ogle/ogle3/transits/ogle56.html.\label{foot:1}}.
We have used these new data to update the ephemeris and the light
curve solution.

\begin{figure}
\figurenum{3}
\hskip 0.1in \includegraphics[scale=1.3]{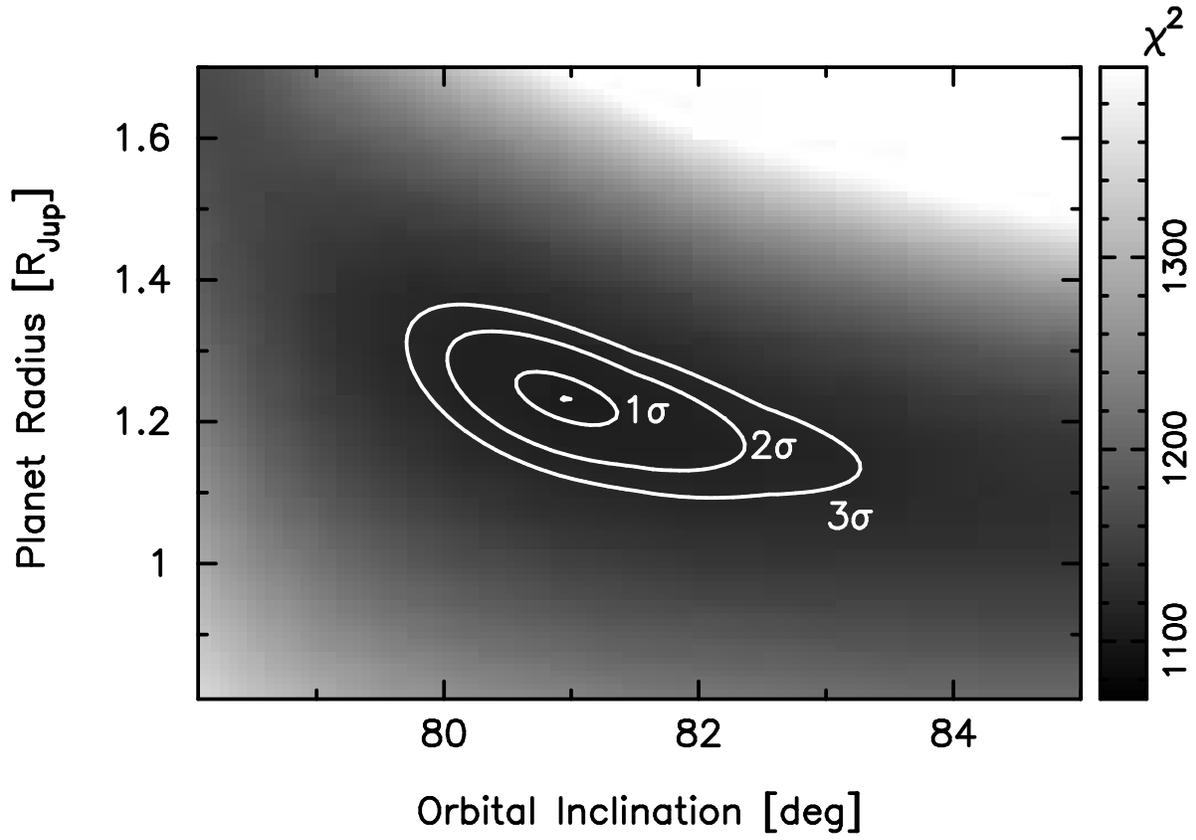}
 \caption{$\chi^2$ surface corresponding to the light curve solution
for OGLE-TR-56, in the plane of planet radius vs.\ orbital
inclination. The number of degrees of freedom in the fit is 1108.
\label{fig:chi2}}
 \end{figure}

\begin{figure}
\figurenum{4}
\hskip -0.1in \includegraphics[scale=0.8]{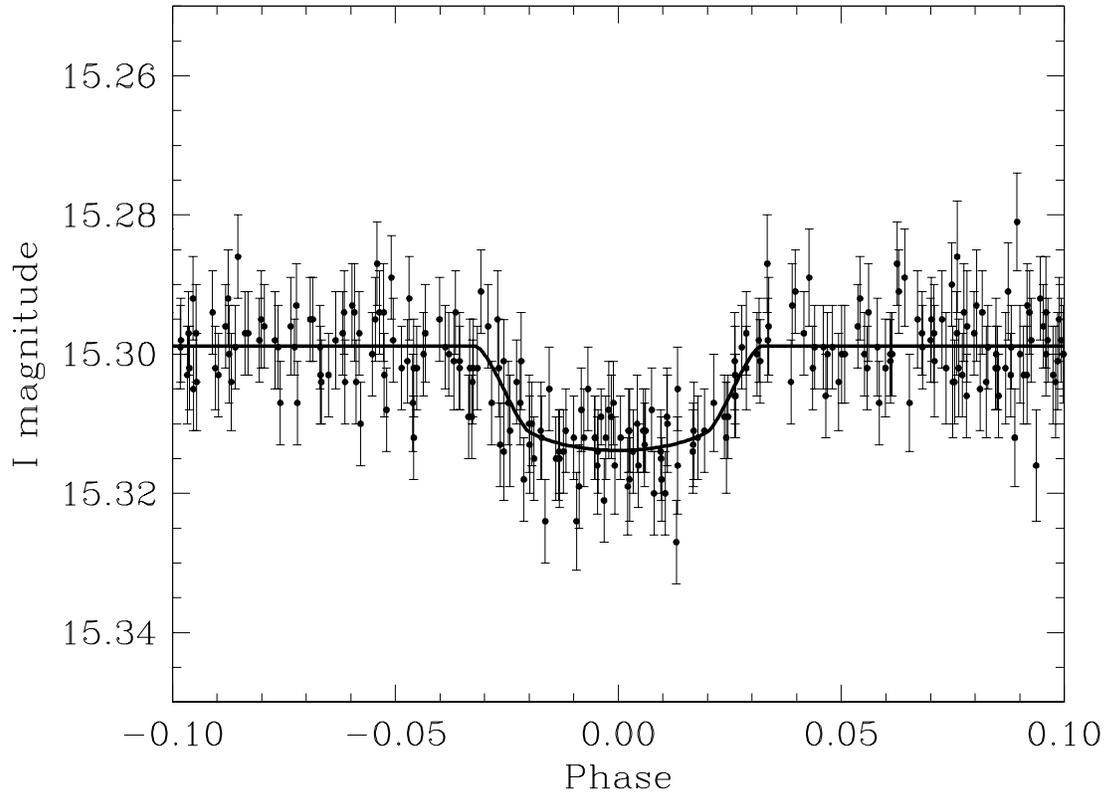}
 \caption{OGLE-III photometry for OGLE-TR-56, and our best fit transit
light curve.\label{fig:lightcurve}}
 \end{figure}

The re-analysis of the transit light curve was carried out with the
tools developed by \cite{Mandel:02}. The stellar parameters (mass and
radius) and the limb darkening coefficient in the $I$ band, $u_I$,
were adopted from \cite{Konacki:03a} and \cite{Sasselov:03}: $M_s =
1.04 \pm 0.05$~M$_{\sun}$, $R_s = 1.10 \pm 0.10$~R$_{\sun}$, $u_I =
0.56 \pm 0.06$. We solved for 5 parameters: the period, transit epoch,
inclination angle, planet radius, and mean magnitude level. The number
of degrees of freedom is 1108.  Figure~\ref{fig:chi2} shows a section
of the $\chi^2$ surface in the vicinity of the minimum, in the plane
of planet radius vs.\ inclination angle. The best fit values are given
in Table~\ref{tab:results}, and the RMS residual of the fit is
0.005~mag. Final errors in the derived parameters include the
contribution from uncertainties in the adopted quantities for the
star, as well as the mass of the planet. These were estimated from
Monte Carlo simulations, and added quadratically to the statistical
errors.  The new ephemeris we derive, $T~{\rm (HJD)} =
2,\!452,\!075.1046(17) + 1.2119189(59)\times n$ (where $n$ is the
number of cycles since the transit epoch), is consistent with that
given in footnote~\ref{foot:1}.  The fit to the OGLE-III photometry is
shown in Figure~\ref{fig:lightcurve}. 
	
\section{Discussion and conclusions}

\begin{figure}
\figurenum{1}
\hskip 0.5in \includegraphics[scale=0.75]{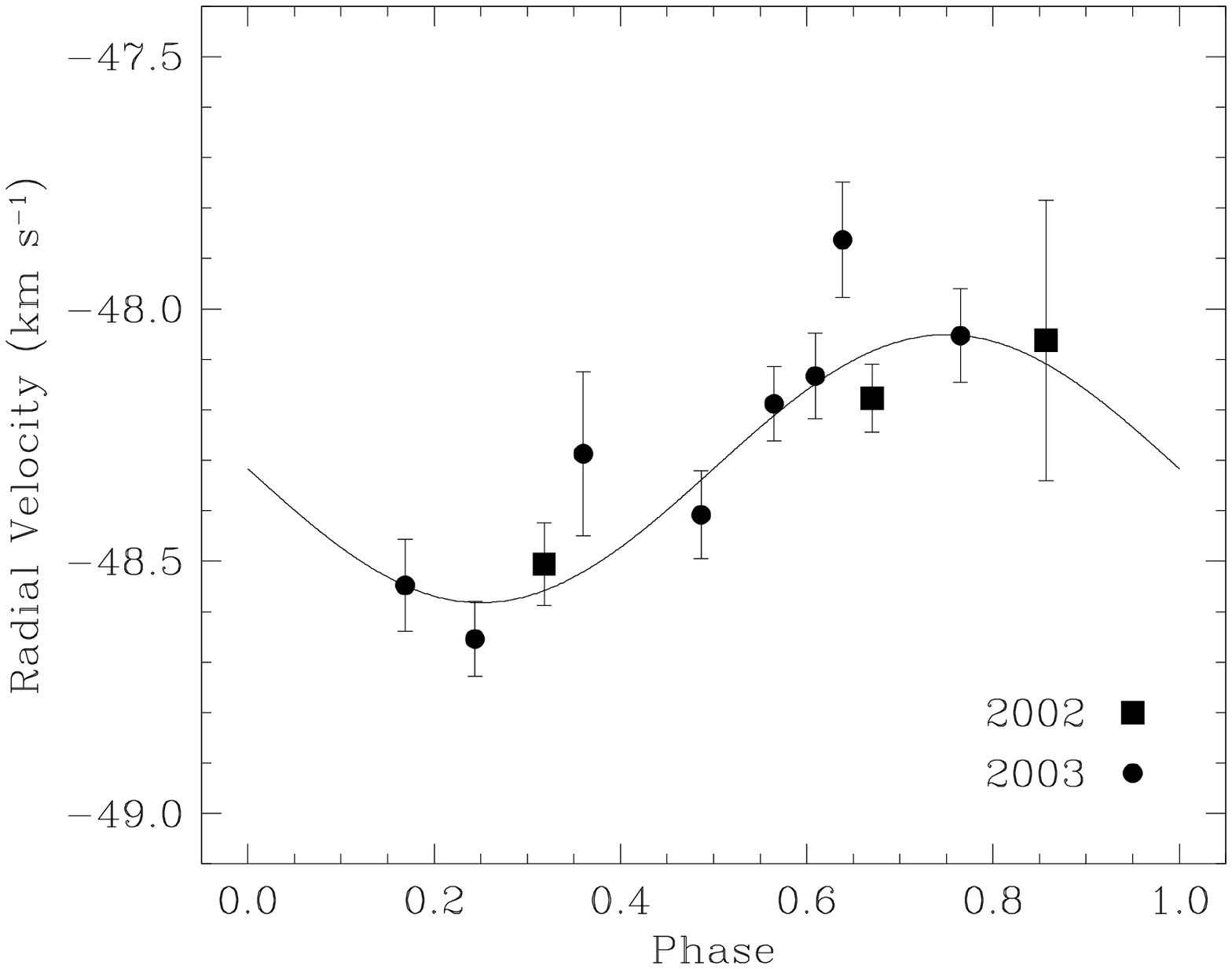}
 \caption{Radial velocity observations and fitted velocity curve for
OGLE-TR-56, as a function of orbital phase (ephemeris from
\S\ref{sec:lightcurve}).\label{fig:orbit}}
 \end{figure}

\begin{figure}
\figurenum{2}
\hskip 0.2in \includegraphics[scale=0.8]{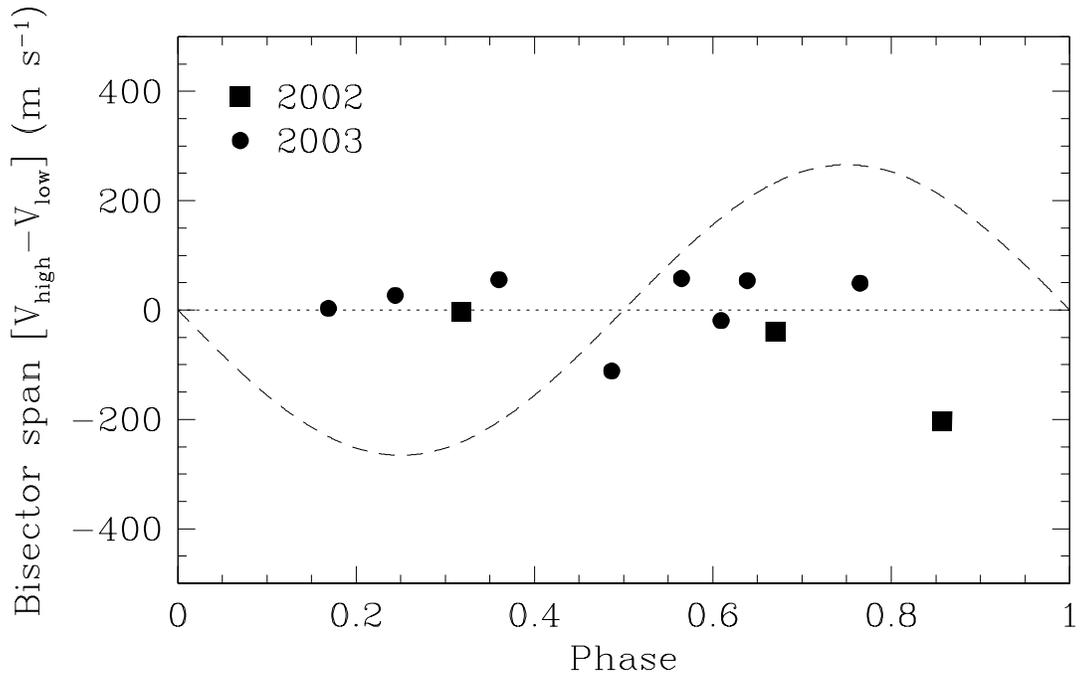}
\vskip -0.5in \caption{Bisector span used as a proxy for line asymmetry for each of
our spectra of OGLE-TR-56, as a function of orbital phase (see text).
Over-plotted for reference is the velocity curve from
Fig.~\ref{fig:orbit}, which shows that there is no correlation of the
asymmetries with phase.\label{fig:bisectors}}
 \end{figure}

Our new radial velocity measurements for OGLE-TR-56 confirm the
variations reported by \cite{Konacki:03a}, and are consistent with the
photometric ephemeris that was held fixed in the orbital solution.
The semi-amplitude we derive using all the data available, $K =
265$~$\ms$, is approximately 60\% larger than the original discovery
estimate ($K = 167$~$\ms$), which was based on only 3 observations
(with two free parameters). The significance of the determination is
now much greater, as can be seen visually in Figure~\ref{fig:orbit},
and the errors are better characterized because of the increased
number of observations.  Consequently, the mass we derive is also
larger: $M_p = 1.45 \pm 0.23$~M$_{\rm Jup}$. The radius, $R_p = 1.23
\pm 0.16$~R$_{\rm Jup}$, is similar to the initial determination. The
reality of the velocity variations is confirmed from the lack of any
significant correlation between the spectral line asymmetries
(bisector spans) and orbital phase. 

\begin{figure}
\figurenum{5}
\hskip 0.1in \includegraphics[scale=0.8]{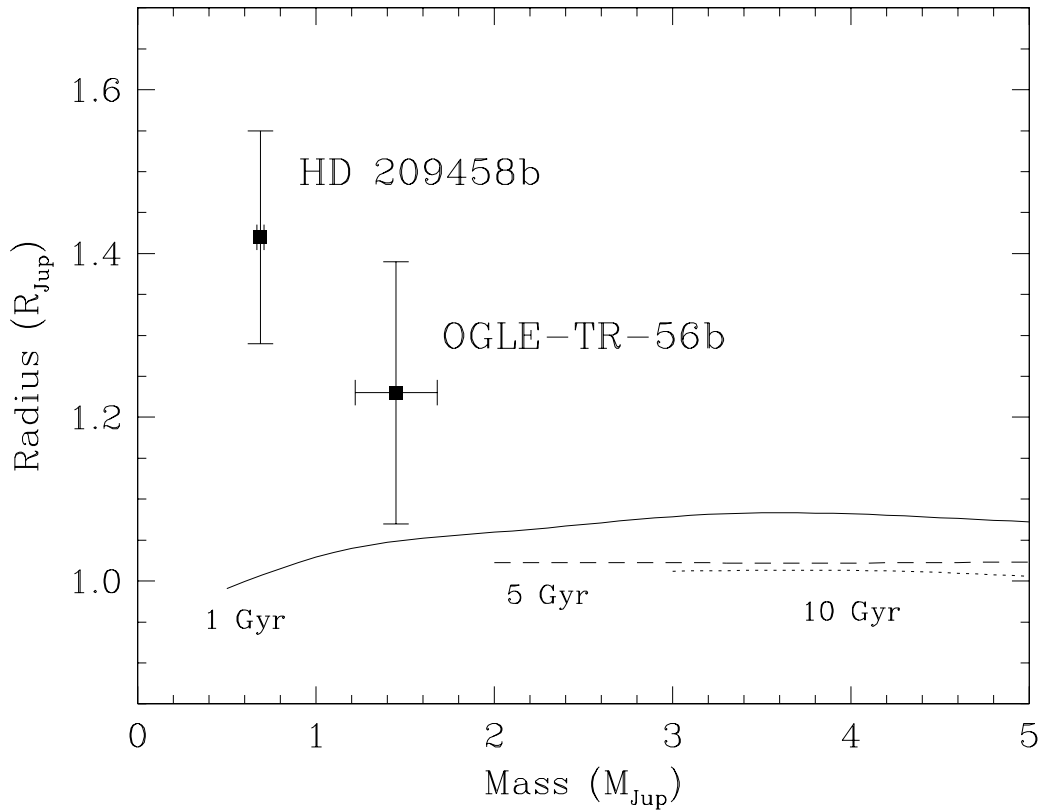}
 \caption{Mass-radius relation by \cite{Baraffe:03} for close-in giant
planets, including the effect of heating by irradiation from the
central star. The observed values for HD~209458b \citep{Cody:02} and
OGLE-TR-56b (this paper) would appear to be inconsistent with these
models at the 3--5 Gyr ages inferred for the two planets. However,
given the uncertainties, OGLE-TR-56 is only moderately inconsistent.
Note also that the models shown (computed specifically for HD~209458b)
have less irradiation than needed for OGLE-TR-56b, which is twice as
close to its parent star.\label{fig:baraffe}}
 \end{figure}

OGLE-TR-56b is roughly twice as massive as HD~209458b, and marginally
smaller \citep[$M_p = 0.69 \pm 0.02$~M$_{\rm Jup}$, $R_p =
1.42^{+0.12}_{-0.13}$~R$_{\rm Jup}$;][]{Cody:02}. Both planets appear
to have radii that are larger than expected from theoretical cooling
models that include a consistent treatment of irradiation by the
parent star (see Figure~\ref{fig:baraffe}). Given the uncertainties
OGLE-TR-56b does not settle the issue, however, and calculations for
the exact conditions of the planet are required
\citep[e.g.,][]{Baraffe:03,Burrows:03}.  Despite the difference in
quality between the OGLE-III light curve for OGLE-TR-56 and the
remarkable HST light curve for HD~209458 \citep{Brown:01}, the error
in our radius determination is not much worse than that of
\cite{Cody:02}. The reason for this is that the dominant contribution
in both cases is the uncertainty in the stellar parameters, which are
at the same level in both cases.  Multicolor HST photometry for both
HD~209458 and OGLE-TR-56 should improve the situation considerably. 
		
\acknowledgments

We are grateful to A.\ Udalski and the OGLE team for their many
generous contributions to this project. We also thank S. Kulkarni for
support and K.\ Stanek for continuous encouragement. The data
presented herein were obtained at the W.\ M.\ Keck Observatory, which
is operated as a scientific partnership among the California Institute
of Technology, the University of California and the National
Aeronautics and Space Administration. The Observatory was made
possible by the generous financial support of the W.\ M.\ Keck
Foundation. G.T.\ acknowledges support from NASA's Kepler
mission. M.K.\ gratefully acknowledges the support of NASA through the
Michelson fellowship program and partial support by the Polish
Committee for Scientific Research, Grant No.~2P03D~001~22., and S.J.\
thanks the Miller Institute for Basic Research in Science at UC
Berkeley for support through a research fellowship. This research has
made use of the SIMBAD database, operated at CDS, Strasbourg, France,
and of NASA's Astrophysics Data System Abstract Service.

\clearpage

%
%

\clearpage

%
%

\clearpage

%
%

\clearpage

%
%

\clearpage

%
%

\clearpage

%
%

\begin{deluxetable}{cccc}
\tablenum{1}
\tablewidth{245pt}
\tablecaption{Radial velocities measurements for OGLE-TR-56, in the
barycentric frame.\label{tab:rvs}}
\tablehead{
\colhead{HJD} & \colhead{} & \colhead{Velocity\tablenotemark{a}} & \colhead{Error\tablenotemark{b}} \\
\colhead{(2,400,000+)} & \colhead{Phase} & \colhead{($\kms$)} & \colhead{($\kms$)} }
\startdata
52480.9239  &  0.8570 & $-$48.062 &   0.278 \\
52481.9095  &  0.6702 & $-$48.177 &   0.067 \\
52483.9068  &  0.3182 & $-$48.506 &   0.082 \\
52853.7474  &  0.4866 & $-$48.408 &   0.087 \\
52853.8960  &  0.6092 & $-$48.133 &   0.085 \\
52854.8062  &  0.3602 & $-$48.287 &   0.163 \\
52855.7863  &  0.1689 & $-$48.548 &   0.091 \\
52855.8772  &  0.2439 & $-$48.654 &   0.074 \\
52863.7802  &  0.7650 & $-$48.053 &   0.093 \\
52864.7497  &  0.5649 & $-$48.188 &   0.074 \\
52864.8389  &  0.6386 & $-$47.863 &   0.114 \\
\enddata
\tablenotetext{a}{Includes a correction of $-192$ $\ms$ to place the
2003 velocities on the same scale as the 2002 measurements (see
text).}
\tablenotetext{b}{Internal errors have been scaled to provide a
reduced $\chi^2$ of unity in the orbital solution (see text).}
 \end{deluxetable}

\clearpage

%
%

\begin{deluxetable}{lc}
\tablenum{2}
\tablewidth{32pc}
\tablecaption{Parameters for OGLE-TR-56b.\label{tab:results}}
\tablehead{
\colhead{\hfil ~~~~~~~~~~~~~~~~~~~~~Parameter~~~~~~~~~~~~~~~~~~~~~~} &  \colhead{Value} }
\startdata
\vspace{2pt}
~~~Orbital period (days)\dotfill                        &  1.2119189~$\pm$~0.0000059   \\
~~~Transit epoch (HJD$-$2,400,000)\dotfill &  52075.1046~$\pm$~0.0017\phm{2222}       \\
~~~Center-of-mass velocity (km~s$^{-1}$)\dotfill       &  $-$48.317~$\pm$~0.045\phm{$-4$} \\
~~~Eccentricity (fixed)\dotfill                 &  0              \\
~~~Velocity semi-amplitude (m~s$^{-1}$)\dotfill        &  265~$\pm$~38\phn              \\
\vspace{10pt}
~~~Inclination angle (deg)\dotfill        &  81.0~$\pm$~2.2\phn  \\
~~~Stellar mass (M$_{\sun}$) (adopted) \dotfill  &  1.04~$\pm$~0.05 \\
~~~Stellar radius (R$_{\sun}$) (adopted) \dotfill  &  1.10~$\pm$~0.10 \\
\vspace{10pt}
~~~Limb darkening coefficient ($I$ band)\dotfill  & 0.56~$\pm$~0.06 \\
~~~{\bf Planet mass (M$_{\rm Jup}$)}\dotfill          &  {\bf 1.45~$\pm$~0.23}      \\
~~~{\bf Planet radius (R$_{\rm Jup}$)}\dotfill        &  {\bf 1.23~$\pm$~0.16}      \\
~~~{\bf Planet density (g~cm$^{-3}$)}\dotfill     &   {\bf 1.0~$\pm$~0.3} \\
~~~Semi-major axis (AU)\dotfill                  &  0.0225~$\pm$~0.0004           \\
\enddata
\end{deluxetable}

\end{document}